\documentstyle[multicol,prd,aps,graphicx]{revtex}
\begin{document}
\draft

\title{ 
Naturally light invisible axion and local $Z_{13}\otimes
Z_3$ symmetries  
} 

\author{Alex G. Dias$^1$, V. Pleitez$^2$ and M. D. Tonasse$^3$}
\address{$^1$ Instituto de F\'\i sica, Universidade de S\~ao Paulo, \\
C. P. 66.318, 05315-970\\ S\~ao Paulo, SP\\ Brazil} 
\address{$^2$ Instituto de F\'\i sica Te\'orica, Universidade Estadual
Paulista,\\ 
Rua Pamplona 145, \\
01405-900 S\~ao Paulo, SP \\Brazil} 
 \address{$^3$Instituto Tecnol\'ogico de Aeron\'autica, Centro T\'ecnico  
Aeroespacial\\ Pra\c ca Marechal do Ar Eduardo Gomes 50, 12228-901 \\
S\~ao Jos\'e 
dos Campos, SP\\ Brazil}  

\date{\today}
\maketitle

\begin{abstract}
We show that by imposing local $Z_{13}\otimes Z_3$ symmetries in an
$SU(2)\otimes U(1)$ electroweak model we can implement an invisible axion 
in such a way that (i) the Peccei-Quinn symmetry is an automatic symmetry of
the classical Lagrangian; and (ii) the axion is protected from
semi classical gravitational effects.
In order to be able to implement such a large discrete symmetry, and at the same
time allow a general mixing in each charge sector, we introduce right-handed
neutrinos and enlarge the scalar sector of the model. The domain wall problem is
briefly considered.   

\end{abstract}

\begin{multicols}{2}
\narrowtext

It has been known for a long time that the complex nature of the QCD
vacuum solves the old $U(1)_A$ problem~\cite{sw,thooft} but also implies
the existence of the $\bar{\theta} F\tilde{F}$ term in the classical
Lagrangian~\cite{theta}. 
This $CP$-nonconserving $\bar{\theta}$ angle (which has also an electroweak
contribution) should be smaller than $10^{-9}$ in order to be consistent with
the measurement of the electric dipole moment of the neutron~\cite{edm}. Why is
$CP$ only weakly violated in QCD? This is the strong $CP$ problem and an elegant
way to solve it is by introducing a chiral symmetry $U(1)_{\rm
PQ}$~\cite{pq}. This Peccei-Quinn (PQ) symmetry implies also the existence of a 
pseudo Goldstone boson---the axion~\cite{axion}. The in\-vi\-si\-ble
axion is almost a singlet (here denoted by $\phi$) under the $SU(2)\otimes U(1)$
gauge symmetry~\cite{singleto}. In addition, since this particle can occur in
the early universe in the form of a Bose condensate which never comes into
thermal equilibrium, it can be a significant component of the cold dark matter
if its mass is of the order of $10^{-5}$ eV~\cite{darkmatter,review}, which
makes it important in its own right. However such a small mass for the axion can
be spoiled by semiclassical gravity effects since
it could induce renormalizable or non-renormalizable effective interactions 
which break explicitly any global symmetry, and in particular the PQ symmetry. 
The most dangerous of these effective o\-pe\-ra\-tors are of the form
$\phi^D$, where $D$ is the mass dimension of the effective operator.
In this case the a\-xi\-on could gain a mass which is greater than the mass
coming from instanton effects and also the $\bar{\theta}$ angle is not
less than $10^{-9}$ in a natural way. This can be  avoided if
the dimension of the effective operators is
$D\stackrel{>}{\sim}12$~\cite{georgi,gravity}. Hence, unless $D$ is high enough,
invisible axion models do not solve the strong CP problem in a natural way. Thus
the invisible (very light) axion must be protected against gravity effects. 

Some years ago Krauss and Wilczek~\cite{kw} pointed out that local
symmetries can masquerade as discrete global symmetries to an observer equipped
with only low-energy probes. We can see this as follows. Consider a
$U(1)$ gauge theory with two complex scalar fields $\eta$ and $\xi$ carrying
charge $Ne$ and $e$, respectively. It is possible, for instance, that $\eta$
undergoes condensation at some very high energy scale $v$, but $\xi$ produces
quanta of relatively small mass and does not condense. Before the condensation
the theory is invariant under the 
transformation  $\eta'(x)=\exp{[iNe\Lambda(x)]}\eta(x)$,  
$\xi'(x)=\exp{[ie\Lambda(x)]}\xi(x)$ and
$A'_\mu(x)=A_\mu(x)-\partial_\mu\Lambda(x)$, here $A_\mu$ is the
$U(1)$ gauge field. The condensate characterized by the vacuum expectation value
$\langle\eta(x)\rangle=v$ in the homogeneous ground state is invariant only when
$\Lambda$ is an integer multiple of $2\pi/Ne$. This residual transformation
still acts nontrivially on $\xi$ but now as a discrete $Z_N$ group. 
The effective theory well below $v$ will be the theory of the single complex
scalar field $\xi$, neither the gauge field nor the scalar $\eta$ will appear in
the effective theory since these fields are very heavy. 
The only implication of the original gauge symmetry for the low energy effective
theory is the absence of interaction terms forbidden by the $Z_N$ symmetry. For
instance if there were more charged scalar fields in the theory,
the discrete symmetry would forbid many couplings that were otherwise possible.
Hence, since no processes like black-hole evaporation or wormhole tunneling
can violate a discrete gauge symmetry it means that local $Z_N$ symmetries
violate the no-hair theorem~\cite{nohair}. These symmetries may be the
consequences of gauge $U(1)$'s symmetries which are embedded in a larger
grand unified theory or from a fifth dimension as it was shown for the case of
$U(1)_Y$ in Ref.~\cite{hill} or even in superstring inspired
models~\cite{gl,dine92,gkn}.

In fact, it has long been known that superstring theories
exhibit PQ-like symmetries. It happens that
in the reduction of a ten-dimensional theory to four dimensions an arbitrary
large number of massless (up to the effect of the coupling $F\tilde{F}$) scalar
four-dimensional fields, which cannot be gauged away,
arise~\cite{witten,green}. On the other hand, in superstring theory,  
also naturally local discrete
symmetries arise~\cite{gl,dine92,gkn,witten}. These facts are 
welcome since superstring theory is considered a good candidate for the
unification of the four interactions. Hence it is reasonable to require
that any global or local symmetry that is introduced in low energy models
can be derived from the superstring theory.  
However, since we already do not know the details of the theory
that des\-cri\-bes quantum gravity, we will apply the ideas above in a
more
conservative scenario: a multi-Higgs extension of the standard model.
 
We show below that in this situation, the axion, as a solution to the
strong CP problem, can be achieved by
introducing appropriated discrete symmetries and 
also the PQ symmetry arises as an automatic or
accidental symmetry of the classical Lagrangian in the sense
that it is not imposed on the Lagrangian but it is just a
consequence of the particle content, renormalizability, and
the gauge and Lorentz invariance of the model~\cite{georgi}. The automatic
PQ symmetry
has been also considered up to now only in the context of grand unified or
supersymmetric theories~\cite{varios} or, recently,
in 3-3-1 models~\cite{pal,iaxion331}. The same can be said with respect to the
axion protection from gravity
effects~\cite{gravity,dine92,gkn,iaxion331}. 

The model we will put forward can be viewed as a way of
stabilizing the Dine-Fischler-Srednicki (DFS) axion~\cite{singleto}. 
As is well known the DFS axion suffers from 
two difficulties: (i) the PQ symmetry has to be imposed by hand and 
(ii) the axion is not
stable from the gravitational effects as discussed before. With respect to
the first problem we expect that the PQ symmetry is not an {\it ad hoc}
symmetry but an accidental one. It could be the consequence of discrete
symmetries $Z_N$ which in addition, if $N$ is large enough, it can stabilize the
axion from the semi-classical gravity effect. In this vain a large $Z_N$
symmetry solves both difficulties of the DFS axion model. The lowest order
effective operator contributing to these parameters have $D=13$ since operators
with dimension up to $D=12$ are forbidden by the local discrete symmetry
$Z_{13}\otimes Z_3$. Moreover, after imposing these discrete symmetries the PQ
chiral symmetry is automatically implemented in a model with $SU(3)_C\otimes
SU(2)_L\otimes U(1)_Y$ gauge symmetry (a similar analysis
for avoiding $B$- and $L$-violating operators in su\-per\-sy\-mme\-tric 
versions of this model was done in Ref.~\cite{lukas}).  

The axion is made invisible if it is almost a singlet
under the gauge group and for this reason we add a singlet $\phi$ which gets a
vacuum expectation value denoted by $v_\phi$~\cite{singleto}. The suppression of
the explicit breakdown of the PQ symmetry by gra\-vi\-ty is obtained by adding a
large $Z_N$ symmetry among the matter multiplets, in such a way that effective
operators like $\phi^{N-1}/M^{(N-1)-4}_{\rm Pl}$, where $M_{\rm Pl}$ is the
Planck mass, are automatically suppressed. At the same time this local discrete
symmetry makes the PQ symmetry an automatic symmetry of the classical
Lagrangian. For instance, a $Z_{13}$ symmetry implies that the first
nonforbidden operator is of dimension thirteen and  the first contribution to
the axion mass square is proportional to $v^{11}_\phi/M^9_{\rm Pl}\approx
10^{-21}\, {\rm eV}^2$ or $10^{-11}m^2_a$, if $m_a\sim\Lambda^2_{\rm
QCD}/v_\phi\approx10^{-5}$ eV is the instanton induced mass (we have used
$ M_{\rm Pl}=10^{19}$ GeV and $v_\phi=10^{12}$ GeV). We see also that the
naturalness of the PQ solution to the $\bar{\theta}$-strong problem is not
spoiled since in this case we have $\theta_{\rm eff}\propto
v^N_\phi/M^{N-4}_{\rm Pl}\Lambda^4_{\rm  QCD}$~\cite{peccei99} and it means also
that $\theta_{\rm eff} \propto 10^{-11}$ for $N=13$. 
  
In order to be able to implement this rather
large symmetry the representation content of the model is augmented: we add
right-handed neutrinos and se\-ve\-ral scalar Higgs multiplets. Presently,
right-handed neutrinos may be a necessary ingredient of any electroweak model
if, as indicated by recent experimental solar~\cite{solarnus} and
atmospheric~\cite{atmosnus} data, neutrinos are massive particles.
On the other hand, each of the extra Higgs scalar multiplets that we will
introduce has already been considered, separately, as viable extensions
of the minimal $SU(2)_L\otimes U(1)_Y$ model~\cite{azee,chengli,hhunter}. 
However, in general the doublets can be coupled to all fermion
doublets, having flavor changing neutral currents if we will. 
In other words, there is not an underlying reason to constraint the
interactions of the multi-Higgs extensions. Below we will show how this
reason arises in an extension of the DFS axion model. 

The representation content is the
following:
$Q_L=(u\,d)^T_L\sim({\bf2},1/3)$, $L_L=(\nu\,l)^T_L\sim({\bf2},-1)$ denote any
quark and lepton doublet; $u_R\sim({\bf1}, 4/3)$, 
$d_R\sim({\bf1},-2/3)$, $l_R\sim({\bf1},-2)$, $\nu_R\sim({\bf1},0)$ are the
right-handed components; and we will assume that each charge
sector gain mass from a different scalar doublets, i.e., $\Phi_u$,
$\Phi_d$, $\Phi_l$, and $\Phi_\nu$ generate Dirac masses for $u$-like, $d$-like
quarks, charged leptons and neutrinos, respectively [all of them of the form
$({\bf2},+1)=(\varphi^+,\,\varphi^0)^T$]. We also add a neutral complex singlet
$\phi\sim({\bf1},0)$ as in \cite{singleto}, a singly charged singlet
$h^+\sim({\bf1},+2)$, as in the Zee's model~\cite{azee} and finally, a triplet
$\vec{{\cal T}}\sim({\bf3},+2)$ that can be written as~\cite{chengli}
\begin{equation}
T\equiv\epsilon\vec{\tau}\cdot \vec{{\cal T}}=\left(
\begin{array}{cc}
\sqrt{2}T^0 & -T^+\\
-T^+&-\sqrt{2}T^{++}
\end{array}
\right),
\label{triplet}
\end{equation}
where $\epsilon=i\tau_2$.

Next, we will impose the following $Z_{13}$ symmetry among those fields: 

\begin{eqnarray}
& & Q\to \omega_5Q,\;u_R \to \omega_3u_R,\;
d_R\to \omega^{-1}_5 d_R,\nonumber \\ 
& & L\to \omega_6 L,\;
\nu_R\to \omega_0\nu_R,\; l_R\to \omega_4 l_R,\nonumber \\ 
& & \Phi_u\to \omega^{-1}_2\Phi_u,\; \Phi_d\to \omega^{-1}_3\Phi_d,\;
\Phi_l\to \omega_2\Phi_l,\nonumber \\
& & \Phi_\nu\to \omega^{-1}_6\Phi_\nu,\;
\phi\to\omega^{-1}_1\phi,\; T\to w^{-1}_4T,\nonumber \\
& &  h^+\to \omega_1h^+,
\label{z13}
\end{eqnarray}
with $\omega_k=e^{2\pi ik/13},\;k=0,1,...,6$.  

With this representation content and the $Z_{13}$ symmetry defined in
Eq.~(\ref{z13}) the allowed Yukawa interactions are given by 
\begin{eqnarray}
-{\cal L}_Y&=&
\overline{Q}_{iL} ( F_{i\alpha }u_{\alpha
R}\tilde{\Phi}_u+\widetilde{F}_{i\alpha }d_{\alpha
R}\Phi_d) + \overline{L_{aL}}(G_{ab}\nu_{bR}\tilde{\Phi}_\nu \nonumber \\ & +&
\widetilde{G}_{ab}l_{bR}\Phi_l)+
f^{ab}\overline{(L_{aiL})^c}L_{bjL}\epsilon_{ij}h^+ + {\rm H.c.}, 
\label{yukawa}
\end{eqnarray}
where $\tilde{\Phi}=\epsilon\Phi^*$, with $i=1,2,3$ and
$a=e,\mu,\tau$; $G,\tilde{G}, F,\tilde{F}$ are arbitrary $3\times3$ matrices and
$f^{ab}$ is a $3\times3$ an\-ti\-sy\-mme\-tric matrix. Notice that we have the
Zee's interaction~\cite{azee} but the interaction
$\overline{(L_{iaL})^c}F_{ab}(\epsilon\vec{\tau}\cdot~\vec{{\cal T}})_{ij}
L_{bLj}$ is not allowed by the $Z_{13}$ symmetry, i.e., this is not an automatic
symmetry of the model. Moreover, the $Z_{13}$ symmetry also implies that there
is no flavor changing neutral currents since there is a doublet for each charge
sector in Eq.~(\ref{yukawa}). The neutrino interactions with the doublet
$\Phi_\nu$ imply a Dirac bare mass; while the singlet $h^+$ induces, through the
radiative corrections, like in the Zee's model, Majorana masses to the
left-handed neutrinos~\cite{koide}.  

The most general scalar potential among the several scalar
multiplets introduced above must be invariant under the gauge
$SU(2)_L\otimes U(1)_Y$ and also under the discrete $Z_{13}$ symmetry. However, 
an extra discrete $Z_3$ symmetry will be added in order to avoid an
interaction in the scalar potential and in Eq.~(\ref{yukawa}). We then have
\end{multicols}
\hspace{-0.5cm}
\rule{8.7cm}{0.1mm}\rule{0.1mm}{2mm}
\widetext
\begin{eqnarray}
V&=&\mu^2_k\Phi^\dagger_k\Phi_k+\mu^2_\phi\phi^*\phi+\mu^2_T{\rm
Tr}(T^\dagger T)+ \mu_h^2h^+h^-+
\lambda^k(\Phi^\dagger_k\Phi_k)^2+
\lambda^{kk'}\vert \Phi^\dagger_k\Phi_{k'}\vert^2+
\lambda^{k\phi}(\Phi^\dagger_k\Phi_k)\phi^*\phi+
\lambda^{kT}_1\Phi^\dagger_k\Phi_k{\rm Tr}(T^\dagger T)
\nonumber \\ &+&
\lambda^{kT}_2(\Phi^\dagger_k T^\dagger)(T\Phi_k)+
\lambda^{kh}\Phi^\dagger_k\Phi_kh^-h^+ +\lambda_1(\phi^*\phi)^2+
\lambda_2({\rm
Tr} T^\dagger T)^2 +
 \lambda_3 {\rm Tr}(T^\dagger T)^2+
\lambda_4{\rm Tr}(T^\dagger T)\vert\phi\vert^2 
+\lambda_5(h^-h^+)^2 \nonumber \\ &+&
\lambda_6{\rm Tr}(T^\dagger T)h^-h^+ 
+[\tilde{\lambda}_1 \Phi^\dagger_d\Phi_l\Phi^\dagger_\nu\Phi_l
+\tilde{\lambda}_2
\tilde{\Phi}^T_d T\tilde{\Phi}_u\phi+
\tilde{\lambda}_3\Phi^{\dagger}_u\Phi_\nu\Phi^\dagger_u\Phi_l
+f_1\Phi^\dagger_u\Phi_d\phi^*+
f_2\tilde{\Phi}^T_u T\tilde{\Phi}_u+
f_3 \tilde{\Phi}^T_l T\tilde{\Phi}_\nu\nonumber \\ &+& {\rm H.c.}], 
\label{potencial}
\end{eqnarray}
\hspace{9.1cm}
\rule{-2mm}{0.1mm}\rule{8.7cm}{0.1mm}
\begin{multicols}{2}
\narrowtext where $k\not=k'$ run over all doublets, i.e., $k,k'=u,d,l,\nu$ and
we have omitted summation symbols. The trilinear 
$\Phi_u\epsilon\Phi_lh^-\phi^*$ is allowed 
by the $Z_{13}$ symmetry but it is forbidden by the $Z_3$ symmetry with
parameters denoted by $\tilde{\omega}_0$, $\tilde{\omega}_1$, and
$\tilde{\omega}^{-1}_1$ if $\Phi_l$ transform with
$\tilde{\omega}^{-1}_1$ and $\Phi_\nu,\nu_R,l_R$ transform with
$\tilde{\omega}_1$ while all the other fields remain invariant i.e.,
transform with $\tilde{\omega}_0$. This $Z_3$ symmetry forbids also a Majorana
mass term $\overline{(\nu_{aR})^c}M_{ab}\nu_{bR}$ in
Eq.~(\ref{yukawa}). It is in this context that
the PQ symmetry is an automatic symmetry of the classical Lagrangian of the
model and the axion mass and the $\bar{\theta}$ angle are protected against
gravitational effects, as discussed in the Introduction.

We have confirmed that the axion is an invisible one. After redefining
the neutral fields as usual $\Phi^0_k=(v_k+{\rm Re}\Phi^0_k+i{\rm
Im}\Phi^0_k)/\sqrt2$,  $T^0=(v_T+{\rm Re}T^0+i{\rm
Im}T^0)/\sqrt2$, and $\phi=(v_\phi+{\rm Re}\phi+i{\rm Im}\phi)/\sqrt2$ there are
only two neutral Goldstone bosons: one which is eaten by the $Z^0$, $G^0$,
and another one which has a small mixture with the $G^0$, the axion which
is almost singlet, i.e., $a\simeq{\rm Im} \phi$. This has been done considering
the full $6\times6$ mass matrix in the pseudoscalar sector.

Besides, the fact that the Peccei-Quinn chiral symmetry
is an automatic symmetry of the model, the discrete $Z_{13}$ symmetry suppresses
also gravitational effects up to operators of dimension twelve. It means that
the contributions to the axion mass and to the $\bar{\theta}$ angle are like
those we have shown above. The assignment of the PQ charges for all the fermion
and scalar fields in the model  are the following (using the notation
$\psi'=e^{-i\alpha X_\psi}\psi$  
with $X_\psi$ the PQ charge of the multiplet $\psi$): 
$X_Q=-3$, $X_{u_R}=3$, $X_{d_R}=3$, $X_L=-6$, $X_{l_R}=-8$, $X_{\nu_R}=4$,
$X_{\Phi_u}=6$, $X_{\Phi_d}=-6$, $X_{\Phi_l}=2$, $X_{\Phi_\nu}=10$, 
$X_T=12$, $X_{h^+}=12$, $X_\phi=-12$.
It is straightforward to verify that the Yukawa
in\-te\-rac\-tions in Eq.~(\ref{yukawa}) and the scalar potential in
Eq.~(\ref{potencial}) are invariant under the PQ transformations given above.
In the present model we have still a
global discrete symmetry, namely $Z_{18}\subset U(1)_{\rm PQ}$. 
Hence, the model has the domain wall problem~\cite{zeldovich} and the usual
mechanism for avoiding it can be applied~\cite{sikivie,dw}.  

Larger $Z_N$ symmetries are possible if we add more fields in any electroweak
model. However, notice that if $N$ is a prime number the axion can transform
under this symmetry with any assignment (but the trivial one); otherwise, we
have to be careful with the way we choose the singlet $\phi$ transforms under
the $Z_N$ symmetry.  

In the present model the axion couples to quarks and leptons (including
neutrinos). Those couplings are proportional to $X_fm_fv^{-1}_\phi$, where
$X_f$ and $m_f$ are the PQ charge and the mass of the fermion $f$.
On the other hand, the axion-photon coupling $c_{a\gamma\gamma}$ is, in general,
proportional to $2N^{-1}_{\rm DW}\sum_fX_fQ_f^2-1.95$, where the sum is over all
generations and the numerical factor comes from the light quarks through a
Fujikawa relation~\cite{kf} and it is the same in the present model; $N_{\rm
DW}$ is the domain wall factor which is equal to 18 in the present model.
We have $c_{a\gamma\gamma}\approx-0.62$,
and not 0.75 as in the invisible axion of
Refs.~\cite{singleto,kaplan,pdg}. The coupling 
of the axion with neutrinos in the present model may have astrophysics and/or
cosmological consequences.  

It is still possible to assign the PQ charge as in Sec.~II of
Ref.~\cite{iaxion331} and the PQ charges are quantized. However, in that case it
appears as an extra Goldstone boson coupling with leptons and photons. Thus, in
order to avoid such an extra Goldstone we have to break softly the $Z_3$
symmetry, say, by adding $\mu^2\Phi_l\Phi_\nu$ in the scalar potential. 

Summarizing, we have obtained a model with an automatic PQ symmetry, with an
invisible axion which is protected against gravitational effects in the context
of a multi-Higgs extension of the standard model. This
is important for three reasons. First, multi-Higgs extensions, 
with or without supersymmetry, are used by the experimentalists as
reference models for searching extra scalar particles~\cite{hhunter,higgs}.
Second, these models are used by phenomenologists to get an appropriate mass
matrix in the lepton and quark sectors. In fact, as we said before, all the
extra fields that we have added to the minimal version of the model have already
been introduced in the literature with different goals; we have just put all of
them together. The necessity of protecting the invisible axion from
gravitational effects, as we have done above, gives a rationale for those
scalar multiplets. Finally, we would like to mention that one of
the issues that remains arbitrary in most multi-Higgs extensions is the
flavor violating neutral currents in the scalar sector. We can implement it or
not at will. However, in our case these effects are automatically avoided since
the Yukawa interactions are already determined for the $Z_{13}$ symmetry. Then,
our model automatically does not have flavor changing neutral currents in all
sectors of the model. 

\acknowledgments 
This work was supported by Funda\c{c}\~ao de Amparo \`a Pesquisa
do Estado de S\~ao Paulo (FAPESP), and partially by Conselho Nacional de 
Desenvolvimento Cientifico e Tecnologico (CNPq).

\end{multicols}


\begin{thebibliography}{99}
\bibitem{sw} S. Weinberg, Phys. Rev. D {\bf11}, 3583 (1975).
\bibitem{thooft} G. t' Hooft,
Phys. Rev. Lett. {\bf37}, 8 (1976); and Phys. Rev. D {\bf 14}, 3432 (1976). 
\bibitem{theta} C. G. Callan, R. Dashen and D. Gross, Phys. Lett. {\bf 63B}, 334
(1976); R. Jackiw and C. Rebbi, Phys. Rev. Lett. {\bf37}, 172 (1976).
\bibitem{edm} V. Baluni, Phys. Rev. D {\bf19}, 2227 (1979); R. Crewther, 
P. Di Vecchia, G. Veneziano and E. Witten, 
Phys. Lett. {\bf 88B}, 123 (1979); {\it ibid}, {\bf91B}, 487(E) (1980). 
\bibitem{pq} R. D. Peccei and H. Quinn, Phys. Rev. Lett. {\bf38}, 1440 (1977);
R. D. Peccei and H. Quinn, Phys. Rev. D {\bf16}, 1791 (1977). 
\bibitem{axion} S. Weinberg, Phys. Rev. Lett. {\bf40}, 223 (1978); F. Wilczek,
{\it ibid}, {\bf40}, 279 (1978).
\bibitem{singleto}  M. Dine, W. Fischler and M. Srednicki, Phys. Lett.
{\bf104B}, 199 (1981). 
\bibitem{darkmatter} M. Srednicki and N. J. C. Spooner, in Particle Data
Group, K. Hagiwara {\it et al.}, Phys. Rev. D {\bf66}, 010001 (2002).
\bibitem{review} Y. E. Kim, Phys. Rep. {\bf150}, 1 (1987); H.-Y. Cheng, {\it
ibid.} {\bf158}, 1 (1998); G. G. Raffelt, Annu. Rev. Nucl. Part. Sci. {\bf49},
163 (1999); Phys. Rep. {\bf198}, 1 (1990); M. S. Turner, {\it ibid.} {\bf197},
67 (1991). 
\bibitem{georgi} H. Georgi, L. J. Hall and M. B. Wise, Nucl. Phys. {\bf B192},
409 (1981); 
\bibitem{gravity} S. Ghigna, M. Lusignoli and M. Roncadelli, Phys. Lett. {\bf
B283}, 278 (1992); R. Holman {\it et al.} Phys. Lett. {\bf B282}, 132 (1992); 
M. Kamionkowski and J. March-Russell, Phys. Lett. {\bf B282}, 137 (1992); S. M.
Barr and D. Seckel, Phys. Rev. D {\bf46}, 539 (1992); R. Holman, T. W. Kephart
and S-J. Rey, Phys. Rev. Lett. {\bf71}, 320 (1993);
R. Kallosh, A. Linde, D. Linde and L. Susskind, Phys. Rev. D {\bf52}, 912
(1995); R. D. Peccei, hep-ph/0009030. 
\bibitem{kw} L. M. Krauss and F. Wilczek, Phys. Rev. Lett. {\bf62}, 1221 (1989).
See also S. Coleman, J. Preskill and F. Wilczek, Nucl. Phys. {\bf B378}, 175
(1992). 
\bibitem{nohair} T. Banks, Nucl. Phys. {\bf B323}, 90 (1988); M. G. Alford, J,
March-Russell and F. Wilczek, Nucl. Phys. {\bf B337}, 695 (1990); J. Preskill
and L. M. Krauss, {\it ibid} {\bf B341}, 50 (1990); M. G. Alford, S. Coleman 
and J. March-Russell, {\it ibid}, {\bf B351}, 735 (1991).
\bibitem{hill} C. T. Hill and A. K. Leibovich, Phys. Rev. D{\bf66}, 075010
(2002); {\it ibid} {\bf 66}, 016006 (2002). 
\bibitem{gl} G. Lazarides, C. Panagiotakopoulos and Q. Shafi, Phys. Rev. Lett.
{\bf56}, 432 (1986); J. A. Casas and G. G. Ross, Phys. Lett. {\bf 192B}, 119
(1987).
\bibitem{dine92} M. Dine, {\sl Problems of Naturalness: Some Lesson From String
Theory}, hep-th/9207045.
\bibitem{gkn} H. Georgi, J. E. Kim and H. P. Nilles,
Phys. Lett. {\bf B437}, 325 (1998).
\bibitem{witten} E. Witten, Nucl. Phys. {\bf B258}, 75 (1985).
\bibitem{green} M. B. Green, J. H. Schwarz and E. Witten, {\sl Superstring
Theory}, (Cambridge University Press, Cambridge, 1987),
Vol. 2; pp. 380-385.  


\bibitem{varios} S. Dimopoulos, P. H. Frampton, H. Georgi and M. B. Wise,
 Phys. Lett. {\bf B117}, 185 (1982); K. Kang and S. Ouvry, Phys. Rev. D
{\bf28}, 2662 (1983); K. S. Soh, Nucl. Phys. {\bf B241}, 129
(1984); Y.J. Park, H-W. Lee and Y. Kim, Phys. Rev. D {\bf30}, 2429
(1984); B. A. Dobrescu, Phys. Rev. D {\bf55},
5826 (1997). 
\bibitem{pal} P. B. Pal, Phys. Rev. D {\bf52}, 1659 (1995).
\bibitem{iaxion331} Alex G. Dias, V. Pleitez and M. D. Tonasse, Phys. Rev.
D{\bf67}, 095008 (2003); Alex G. Dias and V. Pleitez, hep-ph/0308037; A. G.
Dias, C. A. de S. Pires, and P. S. Rodrigues da Silva, hep-ph/0309058.
\bibitem{lukas} E. J. Chun and A. Lukas, Phys. Lett. {\bf B297}, 298
(1992).
\bibitem{peccei99} R. D. Peccei, Nucl. Phys. {\bf B} (Proc. Suppl.) {\bf72}, 3
(1999). 

\bibitem{solarnus}
B. T. Cleveland {\it et al.} (Homestake Collaboration), Nucl. Phys. B (Proc. 
Suppl.) {\bf 38}, 47 (1995); K. S.~Hirata et al., Phys. Rev. D {\bf 44}, 2241
(1991); W. Hampel {\it et al.}
(GALLEX Collaboration), Phys. Lett. {\bf B477}, 127 (1999) J. N. Abdurashitov
{\it at al.} (SAGE Collaboration), Phys. Rev. Lett. {\bf 77}, 4708 (1996);
Q. R. Ahmed {\it et al.} (SNO Collaboration), Phys. Rev. Lett. 
{\bf87}, 071301 (2001). 
\bibitem{atmosnus}
Y. Fukuda et al., \prl {\bf 81}, 1562 (1998); {\it ibid} {\bf 81}, 1158 
(1998); \pl {\bf B436}, 33 (1998); K.S.~Hirata et al., Phys. Lett. {\bf B280},
146 (1992);  R.~Becker-Szendy et al., Phys. Rev. D {\bf 46}, 3720 (1992); 
W. W. M. Allison et al., Phys. Lett. {\bf B 391}, 491 (1997);
Y. Fukuda  et al, Phys. Lett. {\bf B 335}, 237 (1994).
\bibitem{azee} A. Zee, Phys. Lett. {\bf 93B}, 389 (1980).
\bibitem{chengli} T. P. Cheng and L-F. Li, Phys. Rev. D {\bf22}, 2860 (1980).
\bibitem{hhunter} J. F. Gunion, H. E. Haber, G. L. Kane and S. Dawson, {\it 
The Higgs Hunter's Guide} (Addison-Wesley, Reading, MA, 1990).

\bibitem{koide} Y. Kiode, Nucl. Phys. Proc. Suppl. {\bf111}, 294 (2002),
hep-ph/0201250 and references therein.  
\bibitem{zeldovich} Y. B. Zeldovich, I. Y. Kobzarev and L. B. Okun, Sov. Phys.
JETP, {\bf40}, 1 (1975).
\bibitem{sikivie} P. Sikivie, Phys. Rev. Lett. {\bf48}, 1156 (1982).
\bibitem{dw} G. Lazarides and Q. Shafi, Phys. Lett. {\bf 115B}, 21 (1982);  
H. Georgi and M. B. Wise, {\it ibid}, {\bf116B}, 123 (1982).
\bibitem{kf} K. Fujikawa, Phys. Rev. Lett. {\bf42}, 1195 (1979); Phys. Rev. D
{\bf21}, 2848 (1980).
\bibitem{kaplan} D. Kaplan, Nucl. Phys. {\bf B260}, 215 (1985).
\bibitem{pdg} C. Hagmann, K. van Bibber and L. J. Rosenberg, in Particle Data
Group, K. Hagiwara {\it et al}, Phys. Rev. D {\bf66}, 010001-338 (2002). 
\bibitem{higgs} P. Igo-Kemmes, in Particle Data
Group, K. Hagiwara {\it et al}, Phys. Rev. D {\bf66}, 010001-309 (2002).

\end{thebibliography}
\end{document}